\documentclass[10pt]{article}
\usepackage{multicol}
\usepackage{graphicx}
\usepackage{amsmath}
\usepackage[a4paper]{geometry}
\usepackage{hyperref}

\begin{document}

\begin{center}
{\Large \bf Effective (kinetic freeze-out) temperature, transverse
flow velocity, and kinetic freeze-out volume in high energy
collisions}

\vskip1.0cm

Muhammad~Waqas$^{1,2,3,}${\footnote{E-mail:
waqas\_phy313@yahoo.com}},
Fu-Hu~Liu$^{1,2,}${\footnote{Corresponding author. E-mail:
fuhuliu@163.com; fuhuliu@sxu.edu.cn}},
Li-Li~Li$^{1,2,}${\footnote{E-mail: shanxi-lll@qq.com}},
Haidar~Mas$^{\prime}$ud~Alfanda$^{4,}${\footnote{E-mail:
hmasud@mails.ccnu.edu.cn}}
\\

\vskip.25cm

{\small\it $^1$Institute of Theoretical Physics \& State Key
Laboratory of Quantum Optics and Quantum Optics Devices,\\ Shanxi
University, Taiyuan, Shanxi 030006, People's Republic of China

$^2$Collaborative Innovation Center of Extreme Optics, Shanxi
University,\\ Taiyuan, Shanxi 030006, People's Republic of China

$^3$School of Nuclear Science and Technology, University of
Chinese Academy of Sciences,\\ Beijing 100049, People's Republic
of China

$^4$Key Laboratory of Quark and Lepton Physics (MOE) \& Institute
of Particle Physics, Central China Normal University, Wuhan, Hubei
430079, People's Republic of China}
\end{center}

\vskip1.0cm

{\bf Abstract:} The transverse momentum spectra of different types
of particles produced in central and peripheral gold-gold (Au-Au)
and inelastic proton-proton (\(pp\)) collisions at the
Relativistic Heavy Ion Collider (RHIC), as well as in central and
peripheral lead-lead (Pb-Pb) and $pp$ collisions at the Large
Hadron Collider (LHC) are analyzed by the multi-component standard
(Boltzmann-Gibbs, Fermi-Dirac, and Bose-Einstein) distribution.
The obtained results from the standard distribution give an
approximate agreement with the measured experimental data by the
STAR, PHENIX, and ALICE Collaborations. The behavior of the
effective (kinetic freeze-out) temperature, transverse flow
velocity, and kinetic freeze-out volume with the mass for
different particles is obtained, which observes the early kinetic
freeze-out of heavier particles as compared to the lighter
particles. The parameters of emissions of different particles are
observed to be different, which reveals a direct signature of the
mass dependent differential kinetic freeze-out. It is also
observed that the peripheral nucleus-nucleus (\(AA\)) and \(pp\)
collisions at the same center-of-mass energy per nucleon pair are
close in terms of the extracted parameters.
\\



\vskip1.0cm
\begin{multicols}{2}

{\section{Introduction}}

A hot and dense fireball is assumed to form for a brief period of
time ($\sim$ a few fm/$c$) over an extended region after the
initial collisions, which undergoes a collective expansion which
leads to the change in the temperature and volume or density of
the system. Three types of temperatures namely the initial
temperature, chemical freeze-out temperature, and kinetic
freeze-out temperature can be found in the literature, which
describe the excitation degrees of interacting system at the
stages of initial collisions, chemical freeze-out, and kinetic
freeze-out
respectively~\cite{Xu:2014jsa,Chatterjee:2015fua,Chatterjee:2014lfa,
Chatterjee:2014ysa,Rasanen:2016vrr,5a,5b}. There is another type
of temperature, named the effective temperature, which is not a
real temperature and it describes the sum of excitation degree of
the interacting system and the effect of transverse flow at the
stage of kinetic freeze-out.

In principle, the initial stage of collisions happens earlier than
other stages such as the chemical and kinetic freeze-out stages.
Naturally, the initial temperature is the largest and the kinetic
freeze-out temperature is the lowest among the three real
temperatures, while the chemical freeze-out temperature is in
between the initial and kinetic freeze-out temperatures. It does
not get rid of the simultaneity for chemical and kinetic
freeze-outs, which results in the chemical and kinetic freeze-out
temperatures to be the same. The effective temperature is often
larger than the kinetic freeze-out temperature but it is equal to
kinetic freeze-out temperature in case of zero transverse flow
velocity.

To conceive the given nature of the nuclear force and to break the
system into massive
fragments~\cite{Bertsch:1983uv,Moretto:1993ve}, it is a good way
to bring the nucleons in interactions in nucleus-nucleus ($AA$)
collisions at intermediate and high energies and such a process
provokes a liquid-gas type phase transition because lots of
nucleons and other light nuclei are emitted. In $AA$ collisions at
higher energies, a phase transition from hadronic matter to
quark-gluon plasma (QGP) is expected to occur. The volume occupied
by such ejectiles source where the mutual nuclear interactions
become negligible (they only feel the coulombic repulsive force
and are free from the attractive force) is said to be kinetic
freeze-out volume and it has been introduced in various
statistical and thermodynamic
models~\cite{Thakur:2016nfd,Thakur:2016boy}. Like the kinetic
freeze-out temperature, the kinetic freeze-out volume also gives
the information of the co-existence of phase transition. This is
one of the major quantities, which is important in the extraction
of vital observables such as multiplicity, micro canonical heat
capacity, and its negative branch or shape of caloric curves under
the external constraints~\cite{Grosse:1997jk, Borderie:2002R,
DAgostino:1999dod, DAgostino:2001tpv, Chomaz:2000sj}.

It is conceivable that the temperature (volume) of the interacting
system decreases (increases) from the initial state to the latest
kinetic freeze-out stage. During the evolution process, the
transverse flow velocity is existent due to the expansion of the
interacting system. The study of dependence of effective (kinetic
freeze-out) temperature, transverse flow velocity, and kinetic
freeze-out volume on collision energy, event centrality, system
size, and particle rapidity is very significant. We are very
interested in the mentioned quantities in central and peripheral
$AA$ and (inelastic) proton-proton ($pp$) collisions at the
Relativistic Heavy Ion Collider (RHIC) and the Large Hadron
Collider (LHC) over a wide enough energy range in which QGP is
expected to form.

In this paper, we study the dependence of effective (kinetic
freeze-out) temperature, transverse flow velocity, and kinetic
freeze-out volume in central and peripheral gold-gold (Au-Au) and
lead-lead (Pb-Pb) collisions at the RHIC and LHC energies and
compare their peripheral collisions with $pp$ collisions of the
same center-of-mass energy per nucleon pair $\sqrt{s_{NN}}$ (or
the center-of-mass energy $\sqrt{s}$ for $pp$ collisions). Only
62.4 GeV at the RHIC and 5.02 TeV at the LHC are considered as
examples. We present the approach of effective temperature and
kinetic freeze-out volume from the transverse momentum spectra of
the identified particles produced in the mentioned $AA$ and $pp$
collisions. The kinetic freeze-out temperature and transverse flow
velocity are then obtained from particular linear relations.

The remainder of this paper is structured as follows. The
formalism and method are described in Section 2. Results and
discussion are given in Section 3. In Section 4, we summarize our
main observations and conclusions.
\\

{\section{The method and formalism}}

Generally, two main processes of particle production are under
consideration, which includes the soft and hard excitation
processes. The soft excitation process is the strong interactions
among multiple partons, while the hard excitation process is the
more violent collisions between two head-on partons. The soft
excitation process has numerous choices of formalisms, including
but are not limited to the Hagedorn thermal model
(Statistical-bootstrap model)~\cite{R.Hagedorn:1983bf}, the
(multi-)standard distribution~\cite{Cleymans:2012ya}, the Tsallis
and related distributions with various
formalisms~\cite{Zheng:2015mhz}, the blast-wave model with Tsallis
statistics~\cite{Tang:2009ud}, the blast-wave model with Boltzmann
statistics~\cite{Schnedermann:1993ws, Lao:2017skd, 20a,
Abelev:2008ab, Abelev:2009bw}, and other thermodynamic related
models~\cite{Cleymans:2005xv, Andronic:2005yp, Uddin:2011bi,
Adak:2016jtk}. The hard excitation process has very limited choice
of formalisms and can be described by the perturbative quantum
chromodynamics (pQCD)~\cite{Odorico:1982eq, Aamodt:2010my,
Biyajima:2016fpb}.

The experimental data of the transverse momentum ($p_T$) spectrum
of the particles are fitted by using the standard distribution
which is the joint name of Boltzmann-Gibbs, Fermi-Dirac, and
Bose-Einstein distributions which correspond to the factor $S=0$,
$+1$, and $-1$, respectively. The standard distribution at the
mid-rapidity can be demonstrated as~\cite{Cleymans:2012ya}
\begin{align}
f_S(p_T)&= \frac{1}{N}\frac{dN}{dp_T}= \frac{1}{N}\frac{gV'}{(2\pi)^2}p_T\sqrt{p_T^2+m_0^2} \nonumber\\
&\times \bigg[\exp\bigg(\frac{\sqrt{p_T^2+m_0^2}}{T} \bigg)+S
\bigg]^{-1}
\end{align}
in which the chemical potential is neglected, where $N$ is the
experimental number of considered particles, $T$ is the fitted
effective temperature, $V'$ is the fitted kinetic freeze-out
volume (i.e. the interaction volume) of the emission source at the
stage of kinetic freeze-out, $g=3$ (or 2) is the degeneracy factor
for pion and kaon (or proton), and $m_0$ is the rest mass of the
considered particle. As a probability density function, the
integral of Eq. (1) is naturally normalized to 1, i.e., we have
$\int_0^{p_{T\max}}f_S(p_T)dp_T=1$, where $p_{T\max}$ denotes the
maximum $p_T$. At very high energy, the influence of $S=+1$ and
$-1$ can be neglected. Only the Boltzmann-Gibbs distribution is
sufficient to describe the spectra at the RHIC and LHC.

Considering the experimental rapidity range $[y_{\min},y_{\max}]$
around the mid-rapidity, we have the accurate form of Eq. (1) to
be
\begin{align}
f_S(p_T)&= \frac{1}{N} \frac{gV'}{(2\pi)^2}p_T
\int_{y_{\min}}^{y_{\max}} \bigg( \sqrt{p_T^2+m_0^2}
\cosh y -\mu \bigg) \nonumber\\
&\times \bigg[\exp\bigg(\frac{\sqrt{p_T^2+m_0^2}\cosh
y-\mu}{T}\bigg)+S\bigg]^{-1}dy,
\end{align}
where the chemical potential $\mu$ is particle dependent, which
was studied by us recently~\cite{LaoHL}. In high energy
collisions, $\mu_j$ ($j=\pi$, $K$, and $p$) are less than several
MeV, which affects slightly $V'$ comparing with that for
$\mu_j=0$. Then, we may regard $\mu\approx0$ in Eq. (2) at high
energies considered in the present work. In Eqs. (1) and (2), only
$T$ and $V'$ are free parameters.

Usually, we have to use the two-component standard distribution
because single component standard distribution is not enough for
the simultaneous description of very low- ($0\sim0.2-0.3$ GeV/$c$)
and low-$p_T$ ($0.2-0.3\sim2-3$ GeV/$c$ or little more) regions,
which are contributed by the resonance decays and other soft
excitation processes respectively. More than two or
multi-component standard distribution can also be used in some
cases. We have the simplified multi-component ($l$-component)
standard distribution to be
\begin{align}
f_S(p_T)&= \sum\limits_{i=1}^{l}k_i
\frac{1}{N_i} \frac{gV'_i}{(2\pi)^2} p_T\sqrt{p_T^2+m_0^2} \nonumber\\
&\times
\bigg[\exp\bigg(\frac{\sqrt{p_T^2+m_0^2}}{T_i}\bigg)+S\bigg]^{-1},
\end{align}
where $N_i$ and $k_i$ denote respectively the particle number and
fraction contributed by the $i$-th component, and $T_i$ and $V'_i$
denote respectively the effective temperature and kinetic
freeze-out volume corresponding to the $i$-th component.

More accurate form of $l$-component standard distribution can be
written as,
\begin{align}
f_S(p_T)&= \sum\limits_{i=1}^{l}k_i \frac{1}{N_i}
\frac{gV'_i}{(2\pi)^2} p_T \nonumber\\
&\times \int_{y_{\min}}^{y_{\max}}\bigg( \sqrt{p_T^2+m_0^2} \cosh
y -\mu \bigg) \nonumber\\
&\times \bigg[\exp\bigg(\frac{\sqrt{p_T^2+m_0^2} \cosh y -\mu}
{T_i} \bigg) +S\bigg]^{-1}dy.
\end{align}
In Eqs. (3) and (4), only $T_i$, $V'_i$ and $k_i$ ($i\leq l-1$)
are free parameters. Generally, $l=2$ or 3 is enough for
describing the spectra in a not too wide $p_T$ range.

In fact, Eqs. (1) or (2) and (3) or (4) can be used for the
description of $p_T$ spectra and for the extraction of effective
temperature and kinetic freeze-out volume in very low- and
low-$p_T$ regions. The high-$p_T$ ($>3-4$ GeV/$c$) region
contributed by the hard excitation process has to be fitted by the
Hagedorn function~\cite{R.Hagedorn:1983bf} which is an inverse
power law
\begin{align}
f_H(p_T)=\frac{1}{N}\frac{dN}{dp_T}=
Ap_T\bigg(1+\frac{p_T}{p_0}\bigg)^{-n}
\end{align}
which is resulted from the pQCD~\cite{Odorico:1982eq,
Aamodt:2010my, Biyajima:2016fpb}, where $A$ is the normalization
constant, which depends on the free parameters $p_0$ and $n$, and
it results in $\int_0^{p_{T\max}}f_H(p_T)dp_T=1$.

In case of considering the contributions of both the soft and hard
excitation processes, we use the superposition in principle
\begin{align}
f_0({p_T})=kf_S(p_T)+(1-k)f_H(p_T),
\end{align}
where $k$ is the contribution ratio of the soft process and gives
a natural result in $\int_0^{p_{T\max}}f_0(p_T)dp_T=1$. In Eq.
(6), the contribution of soft process is from 0 to $\sim2-3$
GeV/$c$, or even to $\sim3-5$ GeV/$c$ at very high energy, and the
hard component contributes the whole $p_T$ range. There are some
mixtures between the contributions of the two processes in
low-$p_T$ region.

According to the Hagedorn model~\cite{R.Hagedorn:1983bf}, the
contributions of the two processes can be separated completely.
One has another superposition
\begin{align}
f_0(p_T)=A_1 \theta (p_1-p_T)f_S(p_T)+ A_2 \theta
(p_T-p_1)f_H(p_T),
\end{align}
where $\theta(x)$ is the usual step function, and $A_1$ and $A_2$
are the normalization constants which make
$A_1f_S(p_1)=A_2f_H(p_1)$. Equation (7) gives the contribution of
soft process from 0 to $p_1$, while the hard component contributes
from $p_1$ up to the maximum.

In the above two two-component functions (Eqs. (6) and (7)), each
component ($f_S(p_T)$ and $f_H(p_T)$) is a traditional
distribution. In fact, the first component ($f_S(p_T)$) is one of
the Boltzmann-Gibbs, Fermi-Dirac, and Bose-Einstein distributions
if we use a given $S$ such as $S=0$, $+1$, and $-1$. The second
component ($f_H(p_T)$) is the Tsallis-like
distribution~\cite{Zheng:2015mhz} if we let $n=1/(q-1)$ and
$p_0=nT_T$, where $q$ is the entropy index and $T_T$ is the
Tsallis temperature.

We will use only the first component in Eq. (7) due to the reason,
we do not study a very wide $p_T$ range in the present work. In
the case of neglecting the contribution of hard component in
low-$p_T$ region in Eq. (6), the first component in Eq. (6) gives
the same result as the first component in Eq. (7). In fact, Eq.
(4) with $l=2$, that is the two-component standard distribution,
is used in the present work. In addition, considering the
treatment of normalization, the real fitted kinetic freeze-out
volume should be $V_1=N_1V'_1/k_1$ and $V_2=N_2V'_2/(1-k_1)$ which
will be simply used in the following section.

It should be noted that the value of $l$ in the $l$-component
standard distribution has some influences on the free parameters
and then the derived parameters. Generally, $l=1$ is not enough to
fit the particle spectra. In the case of using $l=2$, the
influence of the second component is obvious due to the fact that
the contribution of the first component is not enough to fit the
particle spectra. In the case of using $l=3$, the influence of the
third component is slight due to the fact that the main
contribution is from the first two components, and the
contribution of the third component can be neglected.
\\

{\section{Results and discussion}}

{\subsection{Comparison with the data}}

Figures 1(a) and 1(b) demonstrate the transverse momentum spectra,
$(1/2\pi p_T)d^2N/dp_Tdy$, of the negatively charged particles
$\pi^-$, $K^-$, and $\bar p$ produced in (a) central (0--10\%) and
(b) peripheral (40--80\%) Au-Au collisions at $\sqrt{s_{NN}}=62.4$
GeV. The circles, triangles, and squares represent the
experimental data measured in the mid-rapidity range $-0.5<y<0$ at
the RHIC by the STAR Collaboration~\cite{Shao:2004yb}. The curves
are our fitted results by Eq. (4) with $l=2$. Following each
panel, the results of Data/Fit are presented. The values of the
related parameters ($T_1$, $T_2$, $V_1$, $V_2$, $k_1$, and $N_0$)
along with the $\chi^2$ and number of degree of freedom (ndof) are
given in Table 1. It can be seen that the two-component standard
distribution fits approximately the experimental data measured at
mid-rapidity in Au-Au collisions at the RHIC.

To see the contributions of the two components in Eq. (4) with
$l=2$, as examples, Figs. 1(c) and 1(d) show the contributions of
the first and second components by the dashed and dotted curves
respectively, and the total contribution is given by the solid
curves. Only the results of $\pi^-$ produced in (c) central
(0--10\%) and (d) peripheral (40--80\%) Au-Au collisions at
$\sqrt{s_{NN}}=62.4$ GeV are presented. The circles represented
the same data points as Figs. 1(a) and 1(b). One can see that the
first component contributes mainly in the very low- and low-$p_T$
region, while the second component contributes in a wider region.
There are large overlap region of the two contributions.

\begin{figure*}[!htb]
\begin{center}
\includegraphics[width=16.cm]{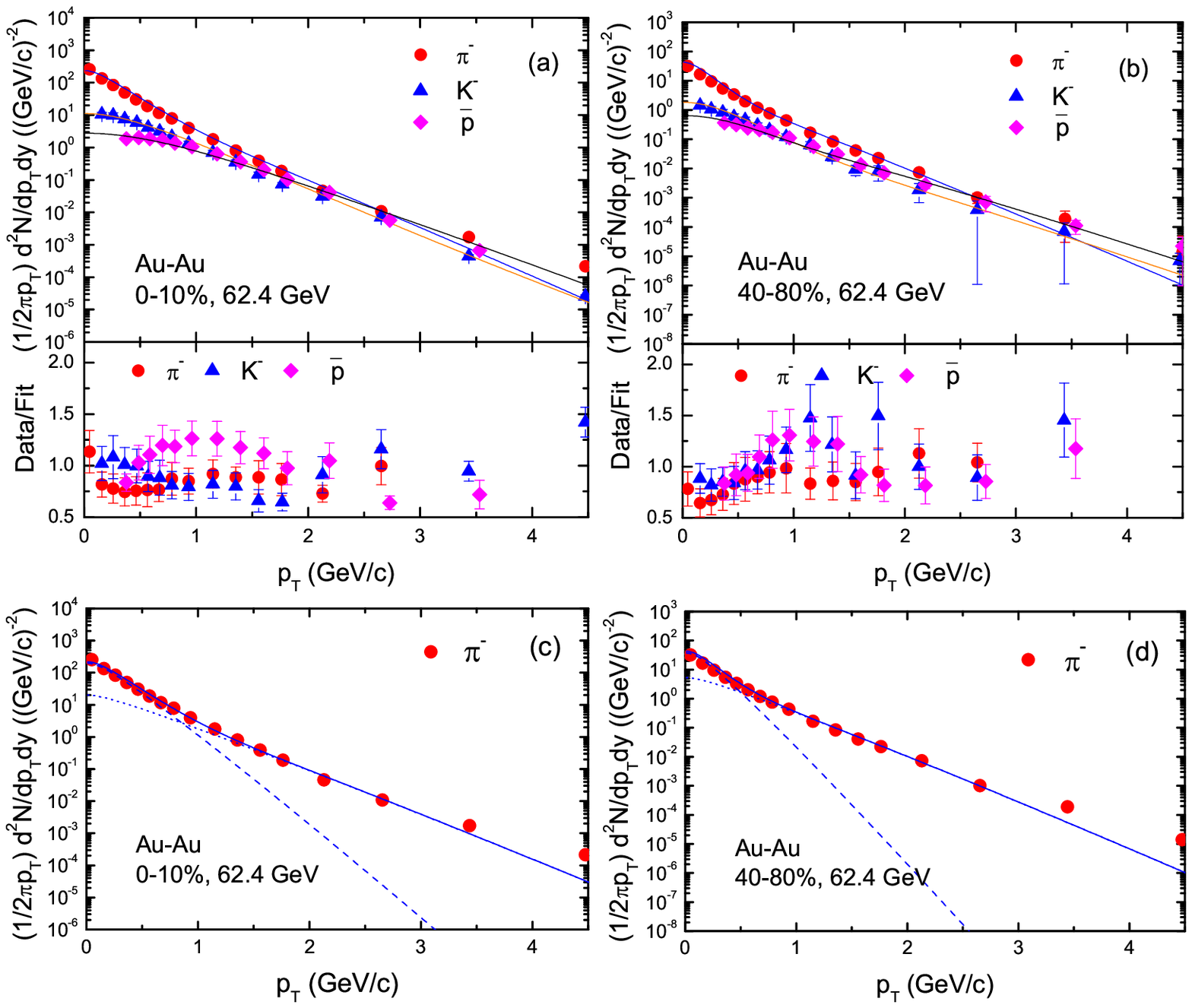}
\end{center}\vskip-0.3cm
Fig. 1. (a)(b) Transverse momentum spectra of $\pi^-$, $K^-$, and
$\bar p$ produced in (a) central (0--10\%) and (b) peripheral
(40--80\%) Au-Au collisions at $\sqrt{s_{NN}}=62.4$ GeV. The
symbols represent the experimental data measured in $-0.5<y<0$ at
the RHIC by the STAR Collaboration~\cite{Shao:2004yb}. The curves
are our fitted results by Eq. (4) with $l=2$. Following each
panel, the results of Data/Fit are presented. (c)(d) As examples,
panels (c) and (d) show the contributions of the first and second
components in Eq. (4) with $l=2$ by the dashed and dotted curves
respectively, and the total contribution is given by the solid
curves. The circles in panels (c) and (d) redisplay those in
panels (a) and (b) respectively.
\end{figure*}

\begin{table*}
{\small Table 1. Values of parameters ($T_1$, $T_2$, $V_1$, $V_2$,
$k_1$, and $N_0$ (for Figs. 1 and 2) or $\sigma_0$ (for Fig. 3)),
$\chi^2$, and ndof corresponding to the solid curves in Figs.
1--3. From the table, we have $k_2=1-k_1$, $T=k_1T_1+k_2T_2$, and
$V=V_1+V_2$. The normalization constants contributed by the first
and second components are $k_1N_0$ (or $k_1\sigma_0$) and $k_2N_0$
(or $k_2\sigma_0$) respectively. \vspace{-.50cm} \scriptsize
\begin{center}
\begin{tabular}{ccccccccccc}\\ \hline\hline
Collisions & Centrality & Particle & $T_1$ (GeV) & $T_2$ (GeV)&
$V_1$ (fm$^3$) & $V_2$ (fm$^3$) & $k_1$ & $N_0$ [$\sigma_0$ (mb)] & $\chi^2$ & ndof \\
\hline
Figure 1  & 0--10\%  & $\pi^-$  & $0.141\pm0.008$ & $0.285\pm0.007$ & $185\pm13$    & $3330\pm270$ & $0.88\pm0.07$ & $0.080\pm0.004$ & 43  & 14\\
Au-Au     &          & $K^-$    & $0.199\pm0.009$ & $0.316\pm0.007$ & $19\pm1$      & $2034\pm162$ & $0.85\pm0.11$ & $0.010\pm0.003$ & 174 & 13\\
62.4 GeV  &          & $\bar p$ & $0.280\pm0.012$ & $0.340\pm0.004$ & $22\pm3$      & $1053\pm100$ & $0.92\pm0.10$ & $0.020\pm0.004$ & 56  & 11\\
\cline{2-11}
          & 40--80\% & $\pi^-$  & $0.070\pm0.006$ & $0.250\pm0.006$ & $32\pm5$      & $127\pm14$   & $0.70\pm0.07$ & $0.025\pm0.050$ & 94  & 14\\
          &          & $K^-$    & $0.239\pm0.008$ & $0.260\pm0.004$ & $2.3\pm0.3$   & $118\pm18$   & $0.89\pm0.09$ & $0.005\pm0.001$ & 40  & 13\\
          &          & $\bar p$ & $0.201\pm0.007$ & $0.302\pm0.005$ & $3.0\pm0.2$   & $69\pm8$     & $0.89\pm0.11$ & $0.009\pm0.001$ & 7   & 11\\
\hline
Figure 2  & 0--5\%   & $\pi^-$  & $0.267\pm0.013$ & $0.624\pm0.005$ & $8943\pm655$  & $4341\pm200$ & $0.93\pm0.12$ & $1.770\pm0.300$ & 425 & 33\\
Pb-Pb     &          & $K^-$    & $0.355\pm0.014$ & $0.465\pm0.006$ & $1820\pm250$  & $5555\pm300$ & $0.94\pm0.12$ & $0.300\pm0.040$ & 776 & 32\\
5.02 TeV  &          & $\bar p$ & $0.459\pm0.014$ & $0.512\pm0.006$ & $381\pm30$    & $5476\pm240$ & $0.94\pm0.10$ & $0.325\pm0.040$ & 748 & 30\\
\cline{2-11}
          & 80--90\% & $\pi^-$  & $0.200\pm0.009$ & $0.407\pm0.004$ & $154\pm8$     & $246\pm50$   & $0.70\pm0.09$ & $0.060\pm0.003$ & 658 & 33\\
          &          & $K^-$    & $0.198\pm0.016$ & $0.420\pm0.005$ & $17\pm2$      & $264\pm45$   & $0.90\pm0.11$ & $0.020\pm0.003$ & 90  & 32\\
          &          & $\bar p$ & $0.302\pm0.018$ & $0.400\pm0.006$ & $4.4\pm0.5$   & $330\pm56$   & $0.92\pm0.13$ & $0.008\pm0.001$ & 296 & 29\\
\hline
Figure 3(a) & $-$    & $\pi^-$  & $0.182\pm0.006$ & $0.275\pm0.005$ & $65\pm8$      & $22\pm4$     & $0.68\pm0.12$ & $0.350\pm0.060$ & 54  & 23\\
$pp$        &        & $K^-$    & $0.160\pm0.007$ & $0.255\pm0.006$ & $5.0\pm0.4$   & $77\pm10$    & $0.88\pm0.15$ & $0.007\pm0.001$ & 4   & 13\\
62.4 GeV    &        & $\bar p$ & $0.235\pm0.008$ & $0.260\pm0.006$ & $1.6\pm0.1$   & $50\pm6$     & $0.95\pm0.10$ & $0.008\pm0.001$ & 126 & 24\\
\hline
Figure 3(b) & $-$    & $\pi^-$  & $0.090\pm0.008$ & $0.370\pm0.005$ & $16\pm2$      & $101\pm13$   & $0.64\pm0.11$ & $0.016\pm0.003$ & 945 & 33\\
$pp$        &        & $K^-$    & $0.850\pm0.013$ & $0.370\pm0.004$ & $0.80\pm0.04$ & $97\pm12$    & $0.87\pm0.11$ & $0.007\pm0.001$ & 666 & 31\\
5.02 TeV    &        & $\bar p$ & $0.539\pm0.010$ & $0.391\pm0.005$ & $1.1\pm0.1$   & $77\pm12$    & $0.90\pm0.13$ & $0.003\pm0.001$ & 496 & 29\\
\hline
\end{tabular}%
\end{center}}
\end{table*}

The transverse momentum spectra, $(1/N_{ev})d^2N/dp_Tdy$, of
$\pi^-$, $K^-$, and $\bar p$ produced in (a) central (0--5\%) and
(b) peripheral (80--90\%) Pb-Pb collisions at $\sqrt{s_{NN}}=5.02$
TeV are shown in Figure 2, where $N_{ev}$ on the vertical axis
denotes the number of events. The experimental data of $\pi^-$,
$K^-$, and $\bar p$ measured in the mid-rapidity range $|y|<0.5$
at the LHC by the ALICE Collaboration~\cite{Yasser Corrales
Morales1,32a} are represented by the circles, triangles, and
squares, respectively. The curves are our results fitted by Eq.
(4) with $l=2$. Following each panel, the results of Data/Fit are
presented. The values of the related parameters along with the
$\chi^2$ and ndof are given in Table 1. One can see that the
two-component standard distribution fits approximately the
experimental data measured at mid-rapidity in Pb-Pb collisions at
the LHC.

\begin{figure*}[!htb]
\vskip.3cm
\begin{center}
\includegraphics[width=16.cm]{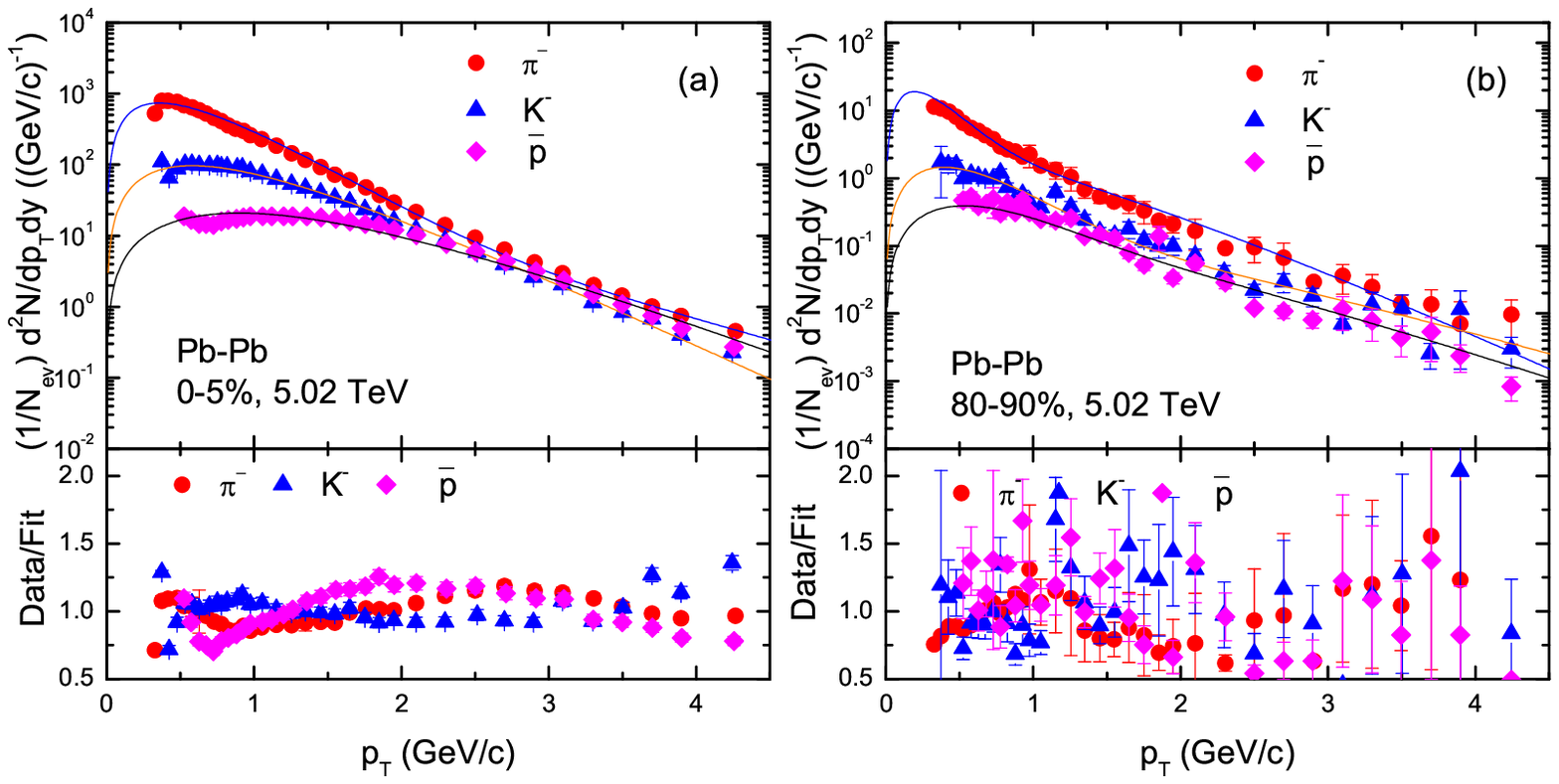}
\end{center}\vskip-0.3cm
Fig. 2. Transverse momentum spectra of $\pi^-$, $K^-$, and $\bar
p$ produced in (a) central (0--5\%) and (b) peripheral (80--90\%)
Pb-Pb collisions at $\sqrt{s_{NN}}=5.02$ TeV. The symbols
represent the experimental data measured in $|y|<0.5$ at the LHC
by the ALICE Collaboration~\cite{Yasser Corrales Morales1,32a}.
The curves are our results fitted by Eq. (4) with $l=2$. Following
each panel, the results of Data/Fit are presented.
\end{figure*}

It seems that Figures 1 and 2 for peripheral collisions show the
worse fits compared to central collisions. This is caused by a
statistical fluctuation and the effect of cold spectator in
peripheral collisions. In the region of cold spectator, particles
are produced by the multiple cascade scattering process which is
different from the thermalization process of particle production
in the region of hot participant. In addition, our fits are done
in all ranges of $p_T<4.5$ GeV/$c$. However, as an alternative
model, the blast-wave fit takes different cuts of $p_T$ for the
analysis of different particles (see for instance
ref.~\cite{Chatterjee:2015fua}). These different cuts affect the
extractions of parameters, in particular for the analysis of the
trends of particles, which is not our expectation.

In the next fits, we also use all ranges of $p_T<4.5$ GeV/$c$.
Figures 3(a) and 3(b) show the transverse momentum spectra,
$Ed^3\sigma/dp^3=(1/2\pi p_T)d^2\sigma/dp_Tdy$, of $\pi^-$, $K^-$,
and $\bar p$ produced in $pp$ collisions at $\sqrt{s}=62.4$ GeV
and 5.02 TeV respectively, where $E$ and $\sigma$ on the vertical
axis denote the energy and cross-section respectively. The symbols
represent the experimental data measured in the mid-pseudorapidity
range $|\eta|<0.35$ by the PHENIX Collaboration~\cite{A.
Adare:2011fb} and in the mid-rapidity range $|y|<0.5$ by the ALICE
Collaborations~\cite{Yasser Corrales Morales1,32a}. The curves are
our results fitted by Eq. (4) with $l=2$. Following each panel,
the results of Data/Fit are presented. The values of the related
parameters ($N_0$ in Figs. 1 and 2 is replaced by $\sigma_0$ in
Fig. 3) along with the $\chi^2$ and ndof are given in Table 1. One
can see that the two-component standard distribution fits
approximately the experimental data measured at
mid-(pseudo)rapidity in $pp$ collisions at the RHIC and LHC.

\begin{figure*}[!htb]
\begin{center}
\includegraphics[width=16.cm]{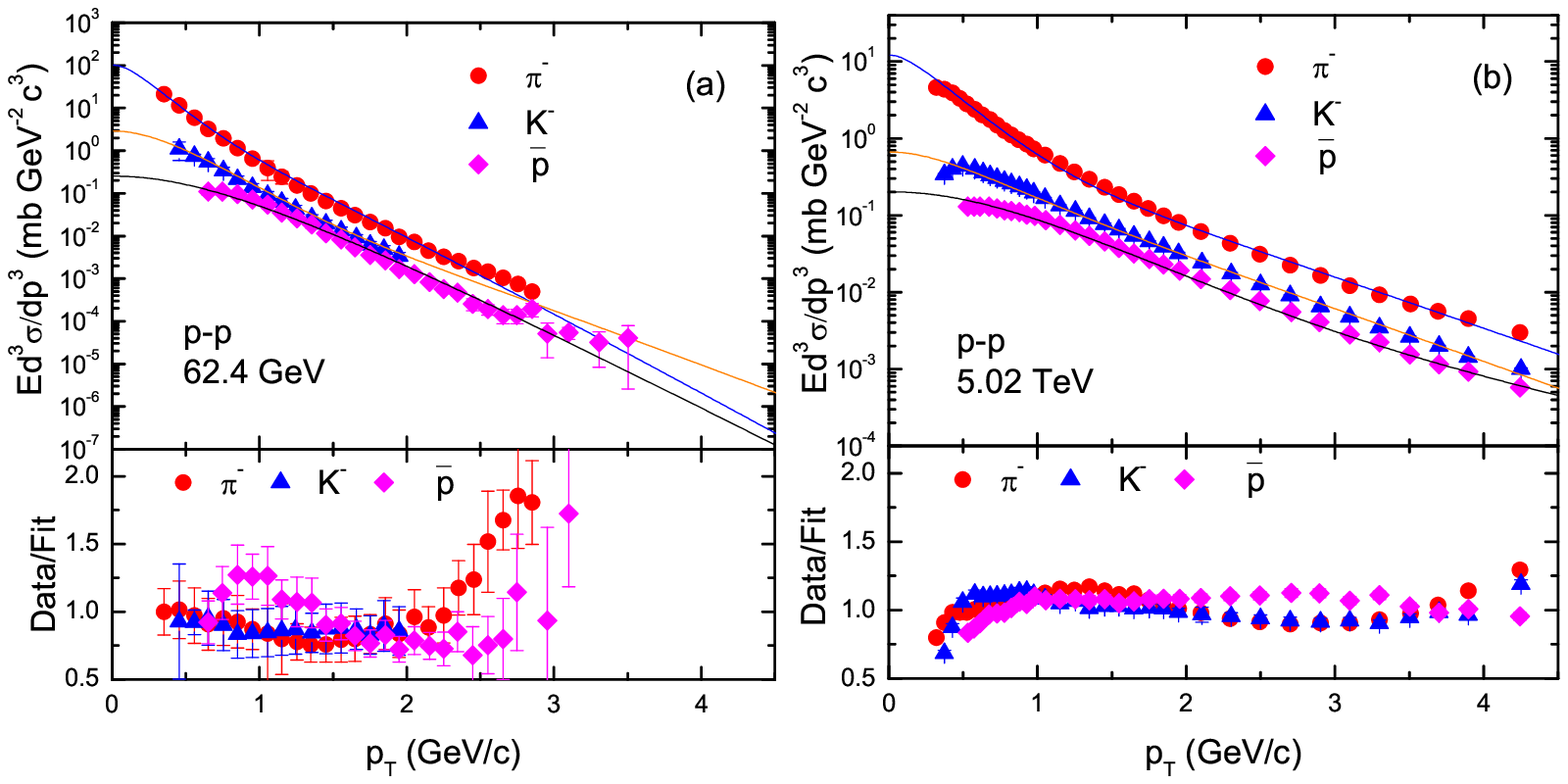}
\end{center}\vskip-0.2cm
Fig. 3. Transverse momentum spectra of $\pi^-$, $K^-$, and $\bar
p$ produced in $pp$ collisions at (a) $\sqrt{s}=62.4$ GeV and (b)
$\sqrt{s}=5.02$ TeV. The symbols represent the experimental data
measured in $|\eta|<0.35$ by the PHENIX Collaboration~\cite{A.
Adare:2011fb} and in $|y|<0.5$ by the ALICE
Collaborations~\cite{Yasser Corrales Morales1,32a}. The curves are
our results fitted by Eq. (4) with $l=2$. Following each panel,
the results of Data/Fit are presented.
\end{figure*}

We would like to point out that the vertical axes of Figs. 1--3
are not the probability density function. We cannot fit them with
Eq. (4) with $l=2$ directly. In fact, we have done a conversion
during our fitting. For Fig. 1, we have used the relation $(1/2\pi
p_T)(d^2N/dp_Tdy)=(1/2\pi p_T)N_0f_S(p_T)/dy$ in the conversion,
where $N_0$ is the normalization constant in terms of particle
number. For Fig. 2, we have used the relation
$d^2N/dp_Tdy=N_0f_S(p_T)/dy$ in the conversion, where $N_{ev}$ on
the vertical axis is neglected because $d^2N/dp_Tdy$ is directly
regarded as the result per event. For Fig. 3, we have used the
relation $Ed^3\sigma/dp^3=(1/2\pi p_T)(d^2\sigma/dp_Tdy)=(1/2\pi
p_T)\sigma_0f_S(p_T)/dy$ in the conversion, where $\sigma_0$ is
the normalization constant in terms of cross section.

From Figures 1--3 and Table 1 one can see that the fit qualities
in some cases are not very well. It should be point out that the
model used in the fits is for soft processes but analyze $p_T$
spectra up to 4.5 GeV/$c$. The highest values of $p_T$ analyzed
contain hard processes which could produce bad fit as indicated by
$\chi^2$ in Table 1 and also in the ratio data to fit of Figs.
1--3, then maybe seems necessary to fit taking into account the
function part corresponding to the hard process. In fact, the hard
process is not necessary for extracting the parameters of soft
process. Although the fits will be better if we also consider the
contribution of hard process, it is not useful for extracting the
parameters considered in the present work. Therefore, we give up
to study the contribution of hard process.
\\

{\subsection{Discussion on the parameters}}

Considering the contributions of the two components, the effective
temperature averaged over the two components is $T=k_1T_1+k_2T_2$
and the kinetic freeze-out volume by adding the two components is
$V=V_1+V_2$. Further, the normalization constants contributed by
the first and second components are $k_1N_0$ and $k_2N_0$
respectively.

For convenience sake, we introduce the average $p_T$ ($\langle
p_T\rangle$) and average moving mass ($\overline{m}$, i.e. average
energy in the source rest frame) here. Considering Eq. (4) only,
we have
\begin{align}
\langle p_T\rangle = \int_0^{p_{T\max}}p_Tf_S(p_T)dp_T.
\end{align}
To obtain $\overline{m}$, we may use the Monte Carlo method. Let
$R_1$ and $R_2$ denote random numbers distributed evenly in [0,1].
A concrete value of $p_T$ which satisfies Eq. (4) can be obtained
by
\begin{align}
\int_0^{p_T}f_S(p_T')dp_T' <R_1< \int_0^{p_T+\delta
p_T}f_S(p_T')dp_T',
\end{align}
where $\delta p_T$ denotes a small shift relative to $p_T$. In the
source rest frame and under the assumption of isotropic emission,
the emission angle $\theta$ of the considered particle obeys
\begin{align}
f_{\theta}(\theta)=\frac{1}{2}\sin\theta
\end{align}
which results in
\begin{align}
\theta=2\arcsin\big(\sqrt{R_2}\big)
\end{align}
in the Monte Carlo method~\cite{37a}. Then,
\begin{align}
m=\sqrt{(p_T/\sin\theta)^2+m_0^2}.
\end{align}
After repetition calculation by many times, we may obtain
$\overline{m}$.

To study the change trends of parameters with the particle mass,
Figures 4(a) and 4(b) show the dependences of $T$ on $m_0$ for
productions of negative charged particles in central and
peripheral (a) Au-Au collisions at 62.4 GeV and (b) Pb-Pb
collisions at 5.02 TeV, while $pp$ collisions at (a) 62.4 GeV and
(b) 5.02 TeV are also studied and compared to peripheral $AA$
collisions of the same energy (per nucleon pair). Correspondingly,
Figures 4(c) and 4(d) show the dependences of $\langle p_T\rangle$
on $\overline{m}$ for the mentioned particles in the considered
collisions. The filled, empty, and half filled symbols represent
central and peripheral $AA$ and $pp$ collisions respectively. The
lines are our linear fittings on the relations. The related linear
fitting parameters are listed in Table 2, though some of them are
not good fitting due to very large $\chi^2$. The intercept in the
linear relation between $T$ and $m_0$ is regarded as the kinetic
freeze-out temperature $T_0$, and the slope in the linear relation
between $\langle p_T\rangle$ and $\overline{m}$ is regarded as the
transverse flow velocity $\beta_T$. That is
$T=am_0+T_0$~\cite{Abelev:2008ab,24a,24b} and $\langle p_T\rangle
=\beta_T \overline{m}+b$, where $a$ and $b$ are free parameters.

It should be noted that the relation
$T=am_0+T_0$~\cite{Abelev:2008ab,24a,24b} is used because the
intercept should be the kinetic freeze-out temperature $T_0$ which
corresponds to the emission of massless particles for which there
is no influence of flow effect. The relation $\langle p_T\rangle
=\beta_T \overline{m}+b$ is used by us in our previous
works~\cite{Lao:2017skd,20a,20b,20c} according to the same
dimension of $\langle p_T\rangle$ and $\beta_T \overline{m}$. The
meanings of the slope $a$ in $T=am_0+T_0$ and the intercept $b$ in
$\langle p_T\rangle =\beta_T \overline{m}+b$ are not clear for us.
Maybe, $am_0$ reflects the effective temperature contributed by
flow effect, and $b$ reflects the average transverse momentum
contributed by the thermal motion.

From Figure 4 and Table 2 one can see that $T$ ($T_0$ or
$\beta_T$) is larger in central $AA$ collisions as compared to
peripheral $AA$ collisions, and peripheral $AA$ collisions are
comparable with the $pp$ collisions at the same $\sqrt{s_{NN}}$
($\sqrt{s}$). The mass dependent or differential kinetic
freeze-out scenario on $T$ is observed, as $T$ increases with the
increase of $m_0$. The present work conforms various mass
dependent or differential kinetic freeze-out
scenarios~\cite{Chatterjee:2015fua, Chatterjee:2014lfa,
Tang:2009ud, Waqas:2018xrz, Waqas:2018tkk}. Because $T_0$
($\beta_T$) is obtained from the linear relation between $T$ and
$m_0$ ($\langle p_T\rangle$ and $\overline{m}$), it seems that
there is no conclusion on the scenario for mass dependent or
independent. However, if we fit firstly $\pi^-$ and $K^-$, and
then including $\bar p$, we can see $T_0$ ($\beta_T$) increases
obviously (decreases slightly) with increasing the mass. Thus, we
observed the mass dependent or differential kinetic freeze-out.

It should be noted that although Figure 4 also shows the
enhancement of $T$ when $m_0$ increases, this has been observed
for many experiments and for the first time was reported by NA44
Collaboration~\cite{35a} as evidence of the flow. That result is
from a fit of $p_T$ to a thermal model for $\pi^-$, $K^-$, and
$\bar p$. This means maybe that the use of the two-component
source model is unnecessary to get the enhancement of $T$ when
$m_0$ increases. In fact, although using a single component source
model can get the similar conclusion, the two-component source
model can describe well the $p_T$ spectra. In addition, including
together the hard component, the model can describe better the
$p_T$ spectra.

The mass dependence of $T$ ($T_0$) and $\beta_T$ is exist because
it reflects the mass dependence of $\langle p_T\rangle$. We do not
think that the mass dependence of $T$ ($T_0$) and $\beta_T$ is a
model dependence, though the values of $T$ ($T_0$) and $\beta_T$
themselves are model dependent. In our fits, we have used the same
$p_T$ range for $\pi^-$, $K^-$, and $\bar p$, while in the
blast-wave fit different $p_T$ ranges are used for the three types
of particles~\cite{Chatterjee:2015fua}. The treatment by the
latter increases the flexibility in the selection of parameters.

Figure 5(a) shows the dependences of kinetic freeze-out volume $V$
on rest mass $m_0$ for productions of negative charged particles
in central and peripheral Au-Au collisions at $\sqrt{s_{NN}}=62.4$
GeV as well as in $pp$ collisions at $\sqrt{s}=62.4$ GeV, while
Figure 5(b) gives the dependences of $V$ on $m_0$ for negative
charged particles produced in central and peripheral Pb-Pb
collisions at $\sqrt{s_{NN}}=5.02$ TeV as well as in $pp$
collisions at $\sqrt{s}=5.02$ TeV. The filled, empty, and half
filled symbols represent the central $AA$, peripheral $AA$ and
$pp$ collisions respectively and they represent the results
weighted different contribution fractions (volumes) in two
components listed in Table 1.

It can be seen from Figure 5 that $V$ in central $AA$ collisions
for all the particles are larger than those in peripheral $AA$
collisions, which shows more participant nucleons and larger
expansion in central $AA$ collisions as compared to peripheral
$AA$ collisions. Meanwhile, $V$ in $pp$ collisions is less than
that in peripheral $AA$ collisions of the same $\sqrt{s_{NN}}$
($\sqrt{s}$), which is caused by the less participant nucleons
(less multiplicity) in $pp$ collisions. It is also observed that
$V$ decreases with the increase of $m_0$. This leads to a volume
dependent or differential freeze-out scenario and indicates
different freeze-out surfaces for different particles, depending
on their masses that show the early freeze-out of heavier
particles as compared to the lighter
particles~\cite{Thakur:2016nfd, Thakur:2016boy}.

\begin{figure*}[!htb]
\begin{center}
\includegraphics[width=16.cm]{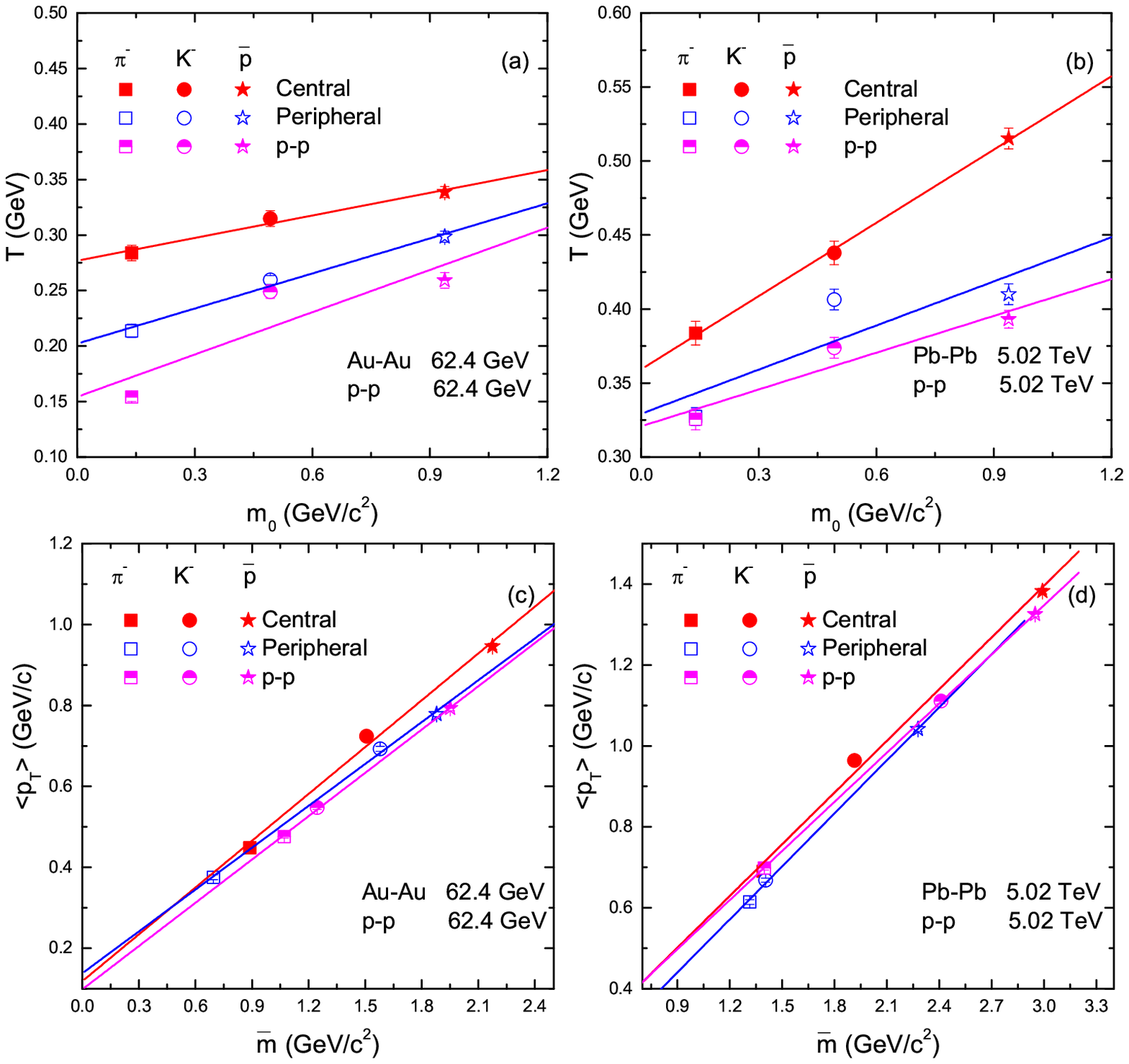}
\end{center}\vskip-0.2cm
Fig. 4. Dependences of (a)(b) $T$ on $m_0$ and (c)(d) $\langle
p_T\rangle$ on $\overline{m}$ for negative charged particles
produced in (a)(c) central and peripheral Au-Au collisions as well
as $pp$ collisions at 62.4 GeV, and in (b)(d) central and
peripheral Pb-Pb collisions as well as $pp$ collisions at 5.02
TeV. The filled, empty, and half filled symbols represent the
parameter values from central and peripheral $AA$ and $pp$
collisions respectively. The lines are linear fits for the
parameter values.
\end{figure*}

\begin{table*}[!htb]
{\small Table 2. Values of slopes, intercepts, and $\chi^2$ in the
linear relations $T=a m_0+T_0$ and $\langle
p_T\rangle=\beta_T\overline{m}+b$, where $T$, $m_0$
($\overline{m}$) and $\langle p_T\rangle$ are in the units of GeV,
GeV/$c^2$ and GeV/$c$ respectively. \vspace{-.20cm}
\begin{center}
\begin{tabular}{ccccccccc}\\ \hline\hline
Figure & Relation & $\sqrt{s_{NN}}$ ($\sqrt{s}$) & Collisions &
$a$ ($c^2$), $\beta_T$ ($c$) & $T_0$ (GeV), $b$ (GeV/$c$)&
$\chi^2$
\\\hline
Figure 4(a) & $T-m_0$ & 62.4 GeV & Central Au-Au    & $0.0679\pm0.006$ & $0.2769\pm0.006$ & 1\\
            &         &          & Peripheral Au-Au & $0.1054\pm0.005$ & $0.2022\pm0.004$ & 1\\
            &         &          & $pp$             & $0.1270\pm0.005$ & $0.1543\pm0.006$ & 40\\
\hline
Figure 4(b) & $T-m_0$ & 5.02 TeV & Central Pb-Pb    & $0.1650\pm0.004$ & $0.3593\pm0.006$ & 1\\
            &         &          & Peripheral Pb-Pb & $0.0994\pm0.005$ & $0.3293\pm0.005$ & 31\\
            &         &          & $pp$             & $0.0829\pm0.005$ & $0.3208\pm0.006$ & 6\\
\hline
Figure 4(c) & $\langle p_T\rangle-\overline{m}$ & 62.4 GeV & Central Au-Au    & $0.3857\pm0.004$ & $0.1186\pm0.006$ & 23\\
            &                                   &          & Peripheral Au-Au & $0.3449\pm0.006$ & $0.1381\pm0.004$ & 5\\
            &                                   &          & $pp$             & $0.3567\pm0.006$ & $0.0983\pm0.005$ & 1\\
\hline
Figure 4(d) & $\langle p_T\rangle-\overline{m}$ & 5.02 TeV & Central Pb-Pb    & $0.4260\pm0.006$ & $0.1178\pm0.005$ & 37\\
            &                                   &          & Peripheral Pb-Pb & $0.4371\pm0.005$ & $0.0465\pm0.004$ & 2\\
            &                                   &          & $pp$             & $0.4048\pm0.006$ & $0.1331\pm0.005$ & 1\\
\hline
\end{tabular}%
\end{center}}
\end{table*}

\begin{figure*}[!htb]
\vskip.3cm
\begin{center}
\includegraphics[width=16.cm]{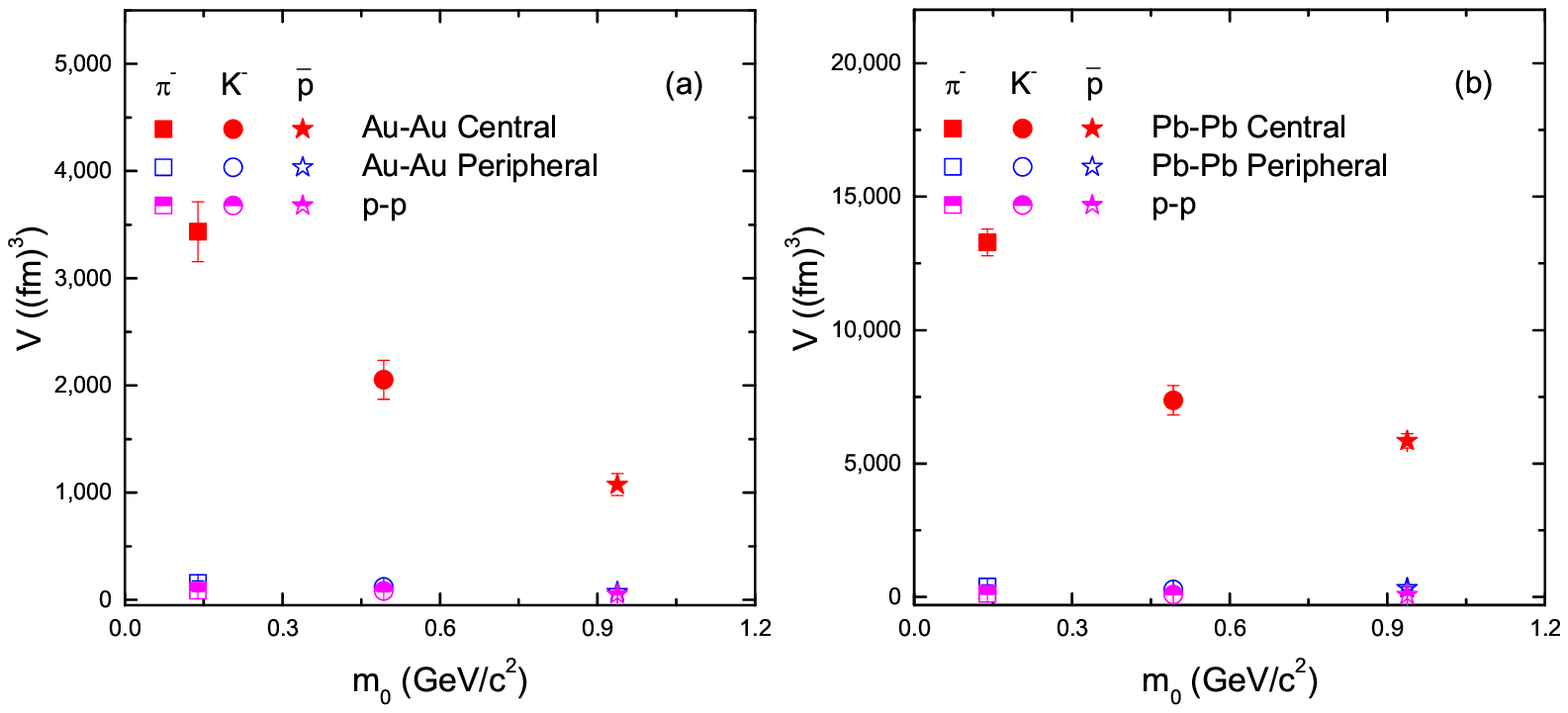}
\end{center}\vskip-0.2cm
Fig. 5. Dependences of $V$ on $m_0$ for negative charged particles
produced in (a) central and peripheral Au-Au collisions as well as
$pp$ collisions at 62.4 GeV, and in (b) central and peripheral
Pb-Pb collisions as well as $pp$ collisions at 5.02 TeV. The
filled, empty, and half filled symbols represent the parameter
values from central and peripheral $AA$ and $pp$ collisions
respectively.
\end{figure*}

\begin{figure*}[!htb]
\vskip.3cm
\begin{center}
\includegraphics[width=16.cm]{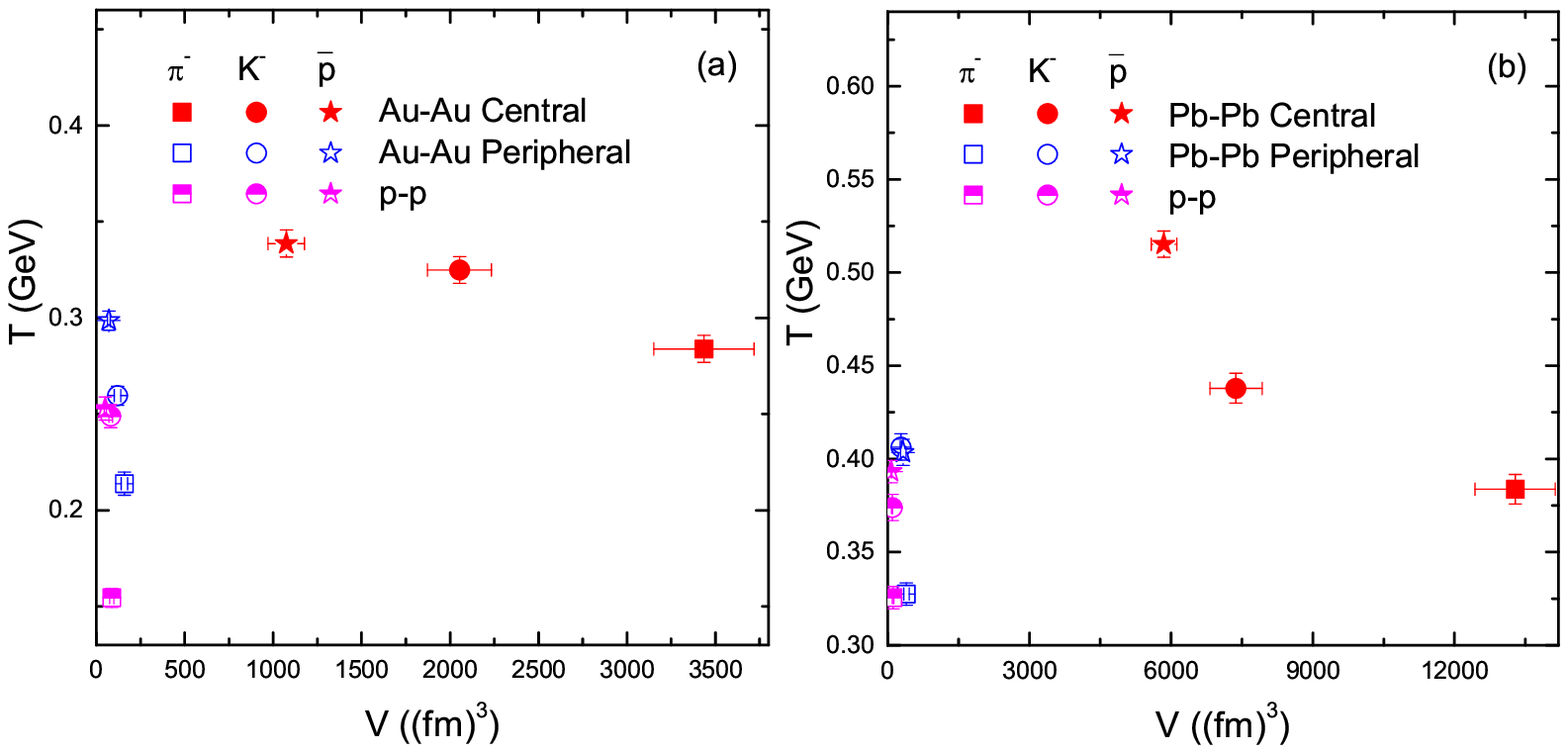}
\end{center}\vskip-0.2cm
Fig. 6. Same as Figure 5, but showing the dependences of $T$ on
$V$.
\end{figure*}

\begin{table*}[!htb]
\vskip.3cm {\small Table 3. Values of $\langle T\rangle$ and
$\langle V\rangle$ in different types of collisions at the RHIC
and LHC. The average values are obtained by different weights due
to different contribution fractions ($V$) of $\pi^-$, $K^-$ and
$\bar p$. \vspace{-.20cm}
\begin{center}
\begin{tabular}{ccccc}\\ \hline\hline
Figure & $\sqrt{s_{NN}}$ ($\sqrt{s}$) & Collisions & $\langle
T\rangle$ (GeV) & $\langle V\rangle$ (fm$^3$) \\ \hline
Figure 1(a) & 62.4 GeV & Central Au-Au    & $0.303\pm0.007$ & $2610\pm218$\\
Figure 1(b) &          & Peripheral Au-Au & $0.247\pm0.007$ & $130\pm17$\\
Figure 3(a) &          & $pp$             & $0.214\pm0.006$ & $77\pm10$\\
\hline
Figure 2(a) & 5.02 TeV & Central Pb-Pb    & $0.478\pm0.009$ & $10002\pm658$\\
Figure 2(b) &          & Peripheral Pb-Pb & $0.374\pm0.008$ & $344\pm54$\\
Figure 3(b) &          & $pp$             & $0.360\pm0.006$ & $100\pm13$\\
\hline
\end{tabular}
\end{center}}
\end{table*}

Figure 6 shows the dependences of $T$ on $V$ for productions of
negative charged particles in (a) central and peripheral Au-Au
collisions as well as in $pp$ collisions at 62.4 GeV, and in (b)
central and peripheral Pb-Pb collisions as well as in $pp$
collisions at 5.02 TeV. The filled, empty, and half filled symbols
represent central and peripheral $AA$ and $pp$ collisions
respectively. One can see that $T$ decreases with the increase of
$V$ in central and peripheral $AA$ and $pp$ collisions. This
result is natural due to the fact that a large $V$ corresponds to
a long kinetic freeze-out time and then a cool system and a low
$T$.

Because of no systematic analysis on the mass dependence of $T_0$
($\beta_T$) being done in the present work, we shall not study the
relation between $T_0$ ($\beta_T$) and $V$ anymore, though we can
predict the trend. As a supplement, our recent work~\cite{35b}
reported the mass dependence (slightly dependence) of $T_0$
($\beta_T$) by the same method as used in the present work, but
using the Tsallis distribution as the ``thermometer". We
understand that with increasing $m_0$ (decreasing $V$), $T_0$
increase obviously, and $\beta_T$ decreases slightly.

From Figures 4--6 one can also see that $T$, $T_0$, $\beta_T$, and
$V$ obtained from collisions at the LHC are larger than those
obtained from collisions at the RHIC. These results are natural
due to more violent collisions happening at higher energy.
However, from the RHIC to LHC, the increase of collision energy is
obviously large, and the increases of $T$, $T_0$, $\beta_T$, and
$V$ are relatively small. This reflects the penetrability of
projectile in the transparent target. In addition, pions
correspond obviously to a larger $V$ than protons in some cases.
This is caused by the fact that, pions have larger $\beta_T$ and
then reach larger space than protons due to less $m_0$ of the
former, at similar momenta for pions and protons at the kinetic
freeze-out. This guess is true due to $V$ being a reflection of
multiplicity, and experimental results indicate an enhancement of
the hadron source as the multiplicity does.

The result that pions correspond to a much larger $V$ than protons
means that the protons stop interacting while pions are still
interacting. In fact, one may think that pions and protons stop
their interacting in different $V$, where large $V$ corresponds to
long interacting time. Because of protons having larger $m_0$ than
pions, protons are leaved behind during the system evolution from
the origin of collisions to radial direction, which is the
behavior of hydrodynamics~\cite{35c}. This results in the volume
dependent freeze-out scenario which shows the early freeze-out of
heavier particles as compared to the lighter
particles~\cite{Thakur:2016nfd, Thakur:2016boy}. Then, pions
correspond to larger interacting volume than protons, at the stage
of kinetic freeze-out.

To study further the dependences of $T$ and $V$ on centrality and
collisions energy, Table 3 contains the values of average $T$
($\langle T\rangle$) and average $V$ ($\langle V\rangle$) for
different types of collisions at the RHIC and LHC. These averages
are obtained by different particle weights due to different
contribution fractions ($V$) of $\pi^-$, $K^-$ and $\bar p$. One
can see clearly that $\langle T\rangle$ and $\langle V\rangle$ at
the LHC are larger than those at the RHIC. Generally, $T$ happens
between $T_{ch}$ and $T_0$. In particular, $T_{ch}$ in central
$AA$ collisions is about 160 MeV, and $T_0$ in central $AA$
collisions is less than 130 MeV~\cite{Cleymans:2005xv,
Andronic:2005yp, A. Andronic:2009fa, A. Andronic:2010bf}. However,
the values of $\langle T\rangle$ in Table 3 are larger because of
Eq. (4) being used. In fact, Eq. (4) contains the contributions of
both thermal motion and flow effect, which can be regarded as a
different ``thermometer" from literature~\cite{Cleymans:2005xv,
Andronic:2005yp, A. Andronic:2009fa, A. Andronic:2010bf,41a},
which results in different $T$ which is beyond the general range
of $[T_{ch},T_0]$.

Even for $T_0$ (the intercept in Table 2 for Figs. 4(a) and 4(b))
obtained from $T=am_0+T_0$, one can see larger values. This is
caused by the different ``thermometers" being used. If one uses
other fit functions~\cite{Abelev:2009bw, Cleymans:2005xv,
Andronic:2005yp, Uddin:2011bi, Adak:2016jtk}, the obtained $T_0$
will be larger or smaller, which depends on the fit function
itself. For $\beta_T$ (the slope in Table 2 for Figs. 4(c) and
4(d)) obtained from $\langle p_T\rangle =\beta_T \overline{m}+b$,
one can see different values from other methods
(``thermometers")~\cite{Tang:2009ud, Schnedermann:1993ws,
Lao:2017skd, Abelev:2008ab}. Anyhow, the relative sizes of $T_0$
($\beta_T$) obtained from the present work for different event
centralities, system sizes, and collision energies are useful and
significative. Generally, $T_0\leq T_{ch}$. However, because
different ``thermometers" are used, it is not simple if we compare
the two temperatures directly.

Although the absolute values of $T$ ($T_0$) and $\beta_T$ obtained
in the present work are possibly inconsistent with other results,
the relative values are clearly worth considering. Similar
situation is true for $V$. The present work shows that $V$ in
central and peripheral Pb-Pb and $pp$ collisions at 5.02 TeV is
also larger than that in central and peripheral Au-Au and $pp$
collisions at 62.4 GeV. This shows strong dependence of parameters
on collision energy. Furthermore, $V$ in central and peripheral
Pb-Pb collisions is larger than that in central and peripheral
Au-Au collisions that also shows somehow parameter dependence on
size of the system, though this dependence can be neglected due to
small difference in the size. The dependence of collision energy
and system size is not discussed here in detail because of the
unavailability of wide range of analysis but it can be focused in
the future work.
\\

{\subsection{Further discussion}}

Before summary and conclusions, we would like to point out that
the method that the related parameters can be extracted from the
$p_T$ spectra of identified particles seems approximately
effective in high energy collisions. In fact, at high energy
(dozens of GeV and above), the particle dependent chemical
potential $\mu$ is less than several MeV which affects indeed less
the parameters. Eqs. (1)--(4) can be used in the present work. We
think that our result on the source volume for $pp$ collisions
being larger than that ($\sim34$ fm$^3$) by the femtoscopy with
two-pion Bose-Einstein correlations~\cite{38} is caused by
different methods.

At intermediate and low energies, the method used here seems
possibly unsuitable due to the fact that the particle dependent
$\mu$ at the kinetic freeze-out is large and unavailable. In
general, the particles of different species develop different
$\mu$ from chemical freeze-out to kinetic freeze-out. This seems
to result in more difficult application of Eqs. (1)--(4) at
intermediate and low energies. In fact, $\mu$ has less influence
on the extraction of source volume due to its less influence on
the data normalization or multiplicity.

As we know, the source volume is proportional to the data
normalization or multiplicity. Although we can obtain the
normalization or multiplicity from a model, the obtained value is
almost independent of model. In other words, the normalization or
multiplicity reflects the data itself, but not the model itself.
Different methods do not affect the source volume obviously due to
the normalization or multiplicity being the most main factor, if
not the only one. In the case of using considerable $\mu$,
neglecting the radial flow and using $T$, there is no obvious
influence on the normalization or multiplicity, then on the source
volume.

In addition, although we use the method of linear relation to
obtain $T_0$ and $\beta_T$ in the present work, we have used the
blast-wave
model~\cite{Tang:2009ud,Schnedermann:1993ws,Abelev:2008ab,
Abelev:2009bw} to obtain the two parameters in our previous
works~\cite{Lao:2017skd, Waqas:2018xrz,Waqas:2018tkk}. Besides, we
could add indirectly the flow velocity in the treatment of
standard distribution~\cite{39}. Because of different
``thermometers" (fit functions) being used in different methods,
the ``measured" temperatures are different in the value, though
the same trend can be observed in the same or similar collisions.
The results obtained from different ``thermometers" can be checked
each other.

In particular, we have obtained the higher temperature, though it
is also the kinetic freeze-out temperature and describes the
excitation degree of emission source at the stage of kinetic
freeze-out. We cannot compare the $T_0$ obtained in the present
work with $T_{ch}$ used in literature directly due to different
``thermometers". We have found that the present work gives the
same trend for main parameters when we compare them with our
previous works~\cite{Lao:2017skd, Waqas:2018xrz,Waqas:2018tkk}
which used the blast-wave
model~\cite{Tang:2009ud,Schnedermann:1993ws,Abelev:2008ab,
Abelev:2009bw}. As it may, the relative size of main parameters in
central and peripheral collisions as well as in $AA$ and $pp$
collisions are the same if we use the standard distribution and
the blast-wave model.

It should be point out that, although we have studied some
parameters at the stage of kinetic freeze-out, the parameters at
the stage of chemical freeze-out are lacking in this paper. In
fact, the parameters at the stage of chemical freeze-out are more
important~\cite{43,44,45,46,47,48} to map the phase diagram in
which $\mu$ is necessary essential factor. Both the $T_{ch}$ and
$\mu$ are the most important parameters at the stage of chemical
freeze-out. In the extensive statistics and/or axiomatic/generic
non-extensive statistics~\cite{43,44,45}, one may discuss the
chemical and/or kinetic freeze-out parameters systematically.

Reference~\cite{46} has tried to advocate a new parametrization
procedure rather than the standard $\chi^2$ procedure with yields,
the authors have constructed the mean value of conserved charges
and have utilized their ratios to extract $T_{ch}$ and $\mu$.
Reference~\cite{47} has evaluated systematic error arising due to
the chosen set of particle ratios and constraints. A centrality
dependent study for the chemical freeze-out parameters~\cite{48}
can be obtained. Meanwhile, with the help of the single-freeze-out
model in the chemical equilibrium framework~\cite{49,50},
Reference~\cite{51} has studied the centrality dependence of
freeze-out temperature fluctuations in high energy $AA$
collisions.

We are very interested to do a uniform study on the chemical and
kinetic freeze-out parameters in future. Meanwhile, the
distribution characteristics of various particles produced in high
energy collisions are very abundant~\cite{52,53,54,55}, and the
methods of modelling analysis are multiple. We hope to study the
spectra of multiplicities, transverse energies, and transverse
momenta of various particles produced in different collisions by a
uniform method, in which the probability density function
contributed by each participant parton is considered carefully.
\\

{\section{Summary and conclusions}}

We summarize here our main observations and conclusions

(a) The main parameters extracted from the transverse momentum
spectra of identified particles produced in central and peripheral
Au-Au collisions at 62.4 GeV and Pb-Pb collisions at 5.02 TeV are
studied. Furthermore, the same analysis is done for $pp$
collisions at both RHIC and LHC energies. The two-component
standard distribution is used, which includes both the very soft
and soft excitation processes. The effective temperature, kinetic
freeze-out temperature, transverse flow velocity, and kinetic
freeze-out volume are larger in central collisions as compared to
peripheral collisions, which shows higher excitation and larger
expansion in central collisions.

(b) The effective temperatures in the central and peripheral Au-Au
(Pb-Pb) collisions at the RHIC (LHC) are observed to increase with
increasing the particle mass, which shows a mass dependent
differential kinetic freeze-out scenario at RHIC and LHC energies.
The kinetic freeze-out temperature is also expected to increase
with increasing the particles mass. The kinetic freeze-out volume
decrease with the increase of particle mass that shows different
values for different particles and it indicates the volume
dependent differential kinetic freeze-out scenario. The transverse
flow velocity is expected to decrease slightly with the increase
of particle mass.

(c) The effective (kinetic freeze-out) temperatures in peripheral
Au-Au and $pp$ collisions at 62.4 GeV as well as in peripheral
Pb-Pb and $pp$ collisions at 5.02 TeV are respectively similar and
have similar trend, which show similar thermodynamic nature of the
parameters in peripheral $AA$ and $pp$ collisions at the same
center-of-mass energy (per nucleon pair). The effective (kinetic
freeze-out) temperatures in both central and peripheral $AA$ and
$pp$ collisions decrease with the increase of the kinetic
freeze-out volume. The transverse flow velocity is expected to
increase slightly with the increase of the kinetic freeze-out
volume in the considered energy range.

(d) The effective (kinetic freeze-out) temperature, transverse
flow velocity, and kinetic freeze-out volume in central and
peripheral $AA$ and $pp$ collisions at the LHC are larger than
those at the RHIC, which shows their dependence on collision
energy. Also, central (peripheral) Pb-Pb collisions give slightly
larger effective (kinetic freeze-out) temperature, transverse flow
velocity, and kinetic freeze-out volume than central (peripheral)
Au-Au collisions. This shows somehow parameter dependence on the
size of the system, which can be neglected for Pb-Pb and Au-Au
collisions due to their small difference in the size.
\\
\\
\\
{\bf Acknowledgments}

This work was supported by the National Natural Science Foundation
of China under Grant Nos. 11575103 and 11947418, the Chinese
Government Scholarship (China Scholarship Council), the Scientific
and Technological Innovation Programs of Higher Education
Institutions in Shanxi (STIP) under Grant No. 201802017, the
Shanxi Provincial Natural Science Foundation under Grant No.
201901D111043, and the Fund for Shanxi ``1331 Project" Key
Subjects Construction.
\\
\\

\end{multicols}
\end{document}